\documentclass[reprint,showpacs,amsmath,amssymb,pra]{revtex4-1}
\usepackage{graphicx}
\usepackage{dcolumn}
\usepackage{bm}

\begin{document}

\title{Propagation of subcycle pulses in a two-level medium: Area-theorem breakdown and pulse shape}

\author{Denis V. Novitsky}
\email{dvnovitsky@tut.by} \affiliation{B. I. Stepanov Institute of
Physics, National Academy of Sciences of Belarus, Nezavisimosti
Avenue 68, BY-220072 Minsk, Belarus}

\date{\today}

\begin{abstract}
We solve the problem of ultrashort pulse propagation in a two-level
medium beyond the rotating-wave (RWA) and slowly-varying-envelope
approximations. The method of solution is based on the
Maxwell--Bloch equations represented in the form that allows one to
switch between RWA and general (non-RWA) cases in the framework of a
single numerical algorithm. Using this method, the effect of a
subcycle pulse (containing less than a single period of field
oscillations) on the two-level medium was analyzed. It is shown that
for such short pulses, the clear breakdown of the area theorem
occurs for the pulses of large enough area. Moreover, deviations
from the area theorem appear to be strongly dependent on the pulse
shape that cannot be observed for longer few-cycle pulses.
\end{abstract}


\maketitle

\section{Introduction}

As the light pulses produced with modern laser techniques become
shorter, so that their duration becomes comparable with the period
of optical field oscillation, the necessity of an adequate
theoretical description of the propagation of such ultrashort pulses
in different systems tends to be more and more obvious. One of the
basic models of an optical medium is a two-level medium, the
fundamental model of a resonantly absorbing medium. The case of
pulses containing only a few cycles of field oscillations implies
that the analysis of the pulse dynamics in the two-level medium
should be carried out beyond the popular and standard approximations
-- the slowly-varying-envelope approximation (SVEA) and the
rotating-wave approximation (RWA). The study of few-cycle pulse
propagation under such generalized conditions (beyond SVEA and RWA)
has been under way since the mid 1990s and has given a number of
important results. For example, the main effects previously known,
such as self-induced transparency (SIT), $2 \pi$ soliton formation,
and $4 \pi$ pulse splitting, were reported to be valid for the
few-cycle pulse though some additional features in the dynamics of
the two-level medium were found as well \cite{Ziolkowski, Kalosha,
Tarasishin}. Among other results one can mention the spectral
transformations due to the intrapulse four-wave mixing
\cite{Kalosha}, the local-field effects on the few-cycle pulse
propagation \cite{Xia, Zhang}, the generation of a single-cycle
soliton in a subwavelength structure consisting of the two-level
medium \cite{Xie}, and the creation of the quasisolitons in a
waveguide-like resonantly absorbing nanostructure \cite{Pusch1}. The
most recent achievements include the effects of the chirp
\cite{Ibanez, Pusch2, Xu} and the so-called counter-rotating terms
in the Bloch equations \cite{Cui} on femtosecond pulse propagation.

In this paper we study the validity of the area theorem for even
shorter (subcycle) pulses. Previously Hughes \cite{Hughes}
discovered the breakdown of the area theorem for the few-cycle
pulses of large area ($2 \pi n$ with $n\geq 3$), while for lower
areas this theorem is still able to predict the profile change of
the pulse accompanied by its splitting \cite{Xiao}. Later Tarasishin
\textit{et al.} \cite{Tarasishin} reported that the half- and
quarter-cycle $2 \pi$ pulses leave some small residual excitation
inside the two-level medium. Here we show that these deviations from
the area theorem are much more noticeable at larger areas of the
incident subcycle pulses and we trace appearance of the theorem
breakdown with a shortening of the pulse. Moreover, these deviations
appear to be strongly dependent on the pulse shape. The effect of
pulse form on the excitation probability of the two-level system is
known for nonresonant excitation (see, for example, the work by
Conover \cite{Conover} and references therein). However, we consider
strictly resonant pulses, their shape being important only for the
number of cycles less than unity.

The structure of the paper is as follows. In Section \ref{eqpars} we
establish the Maxwell--Bloch equations to be numerically solved and
give the main parameters of calculations. In Section \ref{test},
comparing our results with the results known from the literature, we
prove that the method based on the equations stated in the previous
section can be applied to simulate ultrashort pulse propagation
beyond the RWA and SVEA. Section \ref{area} is devoted to the study
of subcycle pulse interaction with the two-level medium, namely to
the issues of the area theorem breakdown and the influence of pulse
shape. Finally, in Section \ref{concl} we give a short conclusion.

\section{\label{eqpars}Main equations and parameters}

Light propagation in the two-level medium beyond the RWA and SVEA is
given by the Maxwell--Bloch equations as follows \cite{Allen,
Kalosha}:
\begin{eqnarray}
\frac{\partial^2 E}{\partial z^2}&-&\frac{1}{c^2} \frac{\partial^2
E}{\partial t^2} = \frac{4 \pi}{c^2} \frac{\partial^2 P}{\partial
t^2}, \label{Max}
\end{eqnarray}
\begin{eqnarray}
\frac{d \rho_{12}}{d t} &=& i \omega_0 \rho_{12} + i
\frac{\mu}{\hbar} E w - \gamma_2 \rho_{12}, \label{polar}
\end{eqnarray}
\begin{eqnarray}
\frac{d w}{d t} &=& -4 \frac{\mu}{\hbar} E \textrm{Im} \rho_{12} -
\gamma_1 (w+1), \label{invers}
\end{eqnarray}
where $E$ is the electric field of a light wave, $\rho_{12}$ the
off-resonant density matrix element (atomic polarization),
$w=\rho_{22}-\rho_{11}$ the inversion (population difference),
$\omega_0$ the frequency of atomic resonance, $\mu$ the dipole
moment of the quantum transition, $\gamma_1$ and $\gamma_2$ the
rates of relaxation of population and polarization, respectively,
$c$ the speed of light, and $\hbar$ the Planck constant. Here the
macroscopic polarization of the two-level medium is $P=2 \mu C
\textrm{Re} \rho_{12}$ with $C$ as the concentration (density) of
two-level atoms. The symbols $\textrm{Re}$ and $\textrm{Im}$ stand
for taking of real and imaginary parts, respectively.

Our aim is to rewrite Eqs. (\ref{Max}) to (\ref{invers}) in such a
manner that they would allow direct comparison of the calculations
conducted with and without the RWA in the framework of a single
numerical algorithm. To reach this aim, we represent the electric
field and atomic polarization as $E= \{ A \exp [i (\omega t - k z)]
+ \textrm{c.c.} \} /2$ and $\rho_{12}=p \exp [i (\omega t - k z)]$,
respectively, but the complex amplitudes $A$ and $p$ are not assumed
to be slowly varying. Here $\omega$ is the central frequency of
radiation, $k=\omega/c$ is the wavenumber, and $\textrm{c.c.}$
stands for complex conjugated term. Introducing dimensionless
arguments $\tau=\omega t$ and $\xi=kz$ and the dimensionless field
amplitude $\Omega=(\mu/\hbar \omega) A$ (normalized Rabi frequency),
we come to the set of equations
\begin{eqnarray}
\frac{\partial^2 \Omega}{\partial \xi^2}&-& \frac{\partial^2
\Omega}{\partial \tau^2}-2 i \frac{\partial \Omega}{\partial \xi}-2
i \frac{\partial \Omega}{\partial
\tau} \nonumber \\
&&=6 \epsilon \left(\frac{\partial^2 p}{\partial \tau^2}+2 i
\frac{\partial p}{\partial \tau}-p\right), \label{Maxdl}
\end{eqnarray}
\begin{eqnarray}
\frac{d p}{d \tau} &=& i \delta p + \frac{i}{2} (\Omega + s \Omega^*
e^{-2 i (\tau-\xi)}) w - \gamma'_2 p, \label{polardl}
\end{eqnarray}
\begin{eqnarray}
\frac{d w}{d \tau} &=& i (\Omega^* p - \Omega p^*) + i s
\left(\Omega p
e^{2 i (\tau-\xi)} - \Omega^* p^* e^{-2 i (\tau-\xi)} \right) \nonumber \\
&&- \gamma'_1 (w+1), \label{inversdl}
\end{eqnarray}
where $\delta=\Delta \omega/\omega=(\omega_0-\omega)/\omega$ is the
frequency detuning, $\gamma'_{1,2}=\gamma_{1,2}/\omega$ are the
normalized relaxation rates, and $\epsilon= \omega_L / \omega=4 \pi
\mu^2 C/3 \hbar \omega$ is the dimensionless parameter of
interaction between light and matter (or normalized Lorentz
frequency). Finally, the auxiliary two-valued coefficient $s$ marks
the situation considered: $s=0$ corresponds to the RWA (absence of
``rapidly rotating'' terms), while $s=1$ is related to the general
case. In this paper we numerically solve Eqs. (\ref{Maxdl}) to
(\ref{inversdl}), so that we have the possibility of switching
between the general (non-RWA) and RWA cases by simply choosing the
appropriate value of the single parameter. The numerical approach is
essentially the same as in our previous publications \cite{Novit,
Novit1, Novit2, Novit3} where the relatively long pulses were
studied in the limit of the RWA (but not the SVEA in the wave
equation). Therefore, we do not discuss the details of the method
and refer the reader to those works.

We adopt the following parameters of the medium and light throughout
the paper: the relaxation rates $\gamma_1=1$ and $\gamma_2=10$
ns$^{-1}$ are large enough so that we are in the regime of coherent
light-matter interaction; the detuning $\delta=0$ (exact resonance);
the light wavelength $\lambda=2 \pi c / \omega=0.83$ $\mu$m; and the
strength of light-matter coupling $\omega_L=10^{11}$ s$^{-1}$ which
is much less than the radiation frequency. These material parameters
mean that one needs to take relatively long thicknesses of the
two-level medium ($L >> \lambda$) to observe the transformations of
the few- or subcycle pulse envelope. Therefore we do not discuss
here the effects of pulse profile changing considered previously for
the relatively stronger coupling conditions ($\omega_L \sim
10^{12}-10^{13}$ s$^{-1}$) \cite{Kalosha, Hughes, Tarasishin}.
Moreover, in the strong coupling limit, when the peak Rabi frequency
$\Omega_0 \omega$ is comparable to or less than the Lorentz
frequency, one needs to take into account the so-called local field
effects \cite{Novit1}. In our research we can neglect them, since
the opposite inequality takes place ($\Omega_0 \omega >> \omega_L$).
The estimation shows that a single-cycle $2 \pi$ soliton in a
collection of two-level atoms with dipole moments $\mu \sim 1$ D
should have the peak amplitude of about $0.4$ GV/cm. The required
concentration in this case is $C \approx 2 \times 10^{19}$
cm$^{-3}$. Obviously, the results of the calculations can be
rescaled for another set of parameters without loss of generality.

In this paper we consider the pulses of two different shapes: the
hyperbolic secant $\Omega=\Omega_p \textrm{sech}(t/t_p)$ and
Gaussian $\Omega=\Omega_p \exp(-t^2/2 t^2_p)$. The duration of the
pulse $t_p$ is defined through the number of cycles $N$ as
$t_p=NT/f$, where $T=\lambda/c$ is the period of electric field
oscillations, and $f$ is the coefficient which depends on the pulse
form and describes its full width at half maximum (FWHM). For the
hyperbolic secant pulse we have $f=2 \textrm{arccosh} \sqrt2$, while
for the Gaussian shape $f=2 \sqrt{\ln 2}$. The peak (normalized)
Rabi frequency $\Omega_p$ is measured in the units of $\Omega_0$
corresponding to the area of $2 \pi$, so that for the hyperbolic
secant pulse one should take $\Omega_0=\lambda/\pi c t_p$ and for
the Gaussian pulse $\Omega_0=\lambda/\sqrt{2 \pi} c t_p$. The
calculational region includes the two-level medium of thickness $L$
surrounded by the vacuum regions of length $0.64$ $\mu$m. The medium
is supposed to be initially in the ground state ($w=-1$).

\section{\label{test}Testing the calculation approach}

\begin{figure}[t!]
\includegraphics[scale=1.1, clip=]{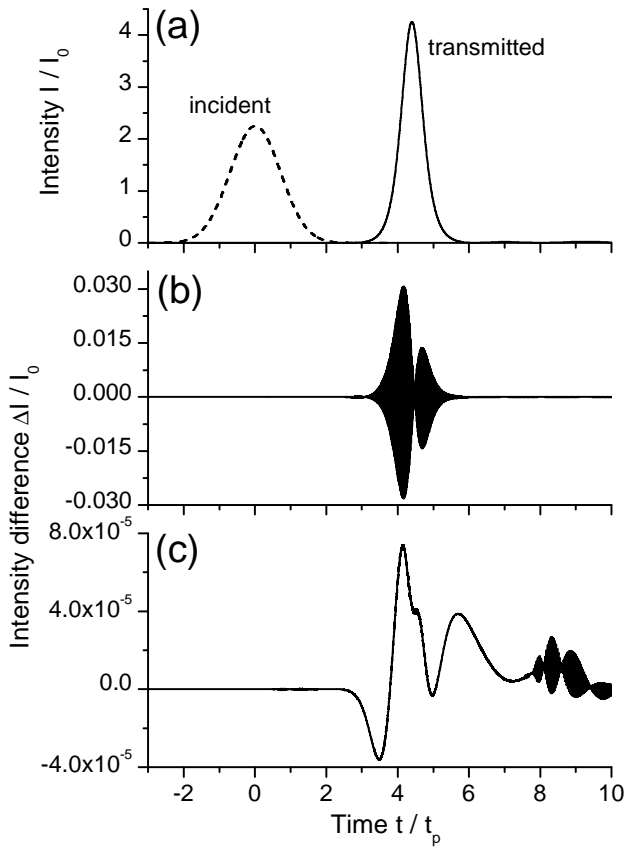}
\caption{\label{fig1} (a) The profile of the Gaussian $50$-cycle $3
\pi$ pulse transmitted through the two-level medium of length $L=100
\lambda$. Calculations were carried out for the general case
($s=1$). (b) The difference between the intensity profiles obtained
for $s=1$ and $s=0$. (c) The difference between the intensity
profiles calculated by two different RWA schemes: the scheme of Eqs.
(\ref{Maxdl}) to (\ref{inversdl}) at $s=0$ and that of Ref.
\cite{Novit}. Intensities are normalized by $I_0=\Omega^2_0$.}
\end{figure}

\begin{figure}[t!]
\includegraphics[scale=1.1, clip=]{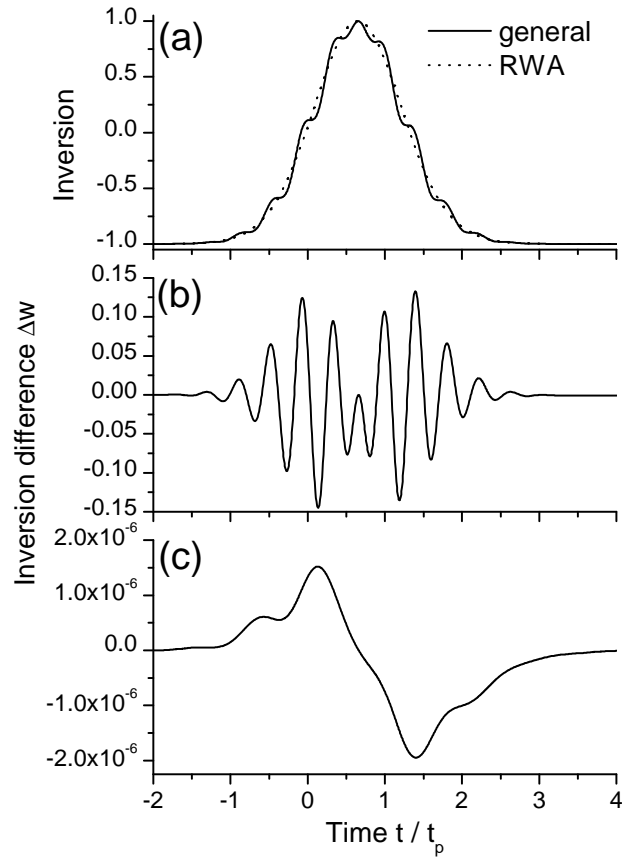}
\caption{\label{fig2} (a) Dynamics of the inversion at the entrance
of the two-level medium excited by the Gaussian two-cycle $2 \pi$
pulse. Calculations were carried out both for the general case
($s=1$) and for the RWA ($s=0$). (b) The difference between the
inversion profiles obtained for $s=1$ and $s=0$. (c) The difference
between the inversion profiles calculated by two different RWA
schemes: the scheme of Eqs. (\ref{Maxdl}) to (\ref{inversdl}) at
$s=0$ and that of Ref. \cite{Novit}.}
\end{figure}

First of all, we need to ascertain that the method based on solving
of Eqs. (\ref{Maxdl}) to (\ref{inversdl}) correctly describes
propagation of the light pulses in the two-level medium. There are
two such tests that are to be considered further.

(i) \textit{The limit of long pulses.} In this situation one expects
that the calculations at $s=1$ and $s=0$ give approximately the same
result. To prove these expectations, we launch the Gaussian $3 \pi$
pulse of $50$ cycles (the duration $t_p \approx 83$ fs) into the
two-level medium of thickness $L=100 \lambda = 83$ $\mu$m. We also
carry out the calculations according to the numerical scheme of Ref.
\cite{Novit} to check the consistency with the RWA case realized in
our previous works \cite{Novit, Novit1, Novit2, Novit3}. Figure
\ref{fig1}(a) shows the intensity profile of such a long pulse
transmitted through the layer of thickness $L=100 \lambda$;
calculations were performed by Eqs. (\ref{Maxdl}) to
(\ref{inversdl}) at $s=1$. One of the main features of such a long
pulse dynamics is seen: the pulse is compressed while forming the
constant-form soliton \cite{Novit2}. Simulations of the RWA scheme
show the agreement with the profile of Fig. \ref{fig1}(a). The plots
in Figs. \ref{fig1}(b) and \ref{fig1}(c) demonstrate the accuracy of
this agreement: the difference between the intensity profiles
obtained for $s=1$ and $s=0$ is as low as several hundredth of $I_0$
(the unit of intensity equal to $\Omega^2_0$), while the discrepancy
between the calculations by two RWA schemes does not exceed $10^{-4}
I_0$, respectively. This proves the correctness of our approach in
the long pulse limit.

(ii) \textit{Behavior of inversion in the case of a few-cycle
pulse.} Since the shape of the few-cycle pulse varies too slowly as
it propagates, our second test deals with the time variation of the
inversion $w$ at the entrance of the two-level medium. To excite the
medium, the two-cycle $2 \pi$ Gaussian pulse is used (its duration
is about $3.32$ fs). The results of calculations are demonstrated in
Fig. \ref{fig2}. The difference between $s=1$ and $s=0$ cases is
clearly seen and reaches values as large as $0.15$. The inversion
profile calculated without the RWA shows also the feature
characteristic for the few-cycle pulse propagation -- the step-like
flattenings corresponding to the extremes of the time derivative of
the electric field \cite{Ziolkowski}. Finally, Fig. \ref{fig2}(c) is
the evidence of the precise correspondence between the calculation
at $s=0$ and the RWA calculations according to the previously used
approach. Thus, we conclude that our numerical method based on Eqs.
(\ref{Maxdl}) to (\ref{inversdl}) allows one to reproduce the main
peculiarities of long- and short-pulse propagation discovered
previously and can be used to study the subcycle pulses beyond the
RWA.

\section{\label{area}Results on subcycle pulses}

\begin{figure}[t!]
\includegraphics[scale=0.9, clip=]{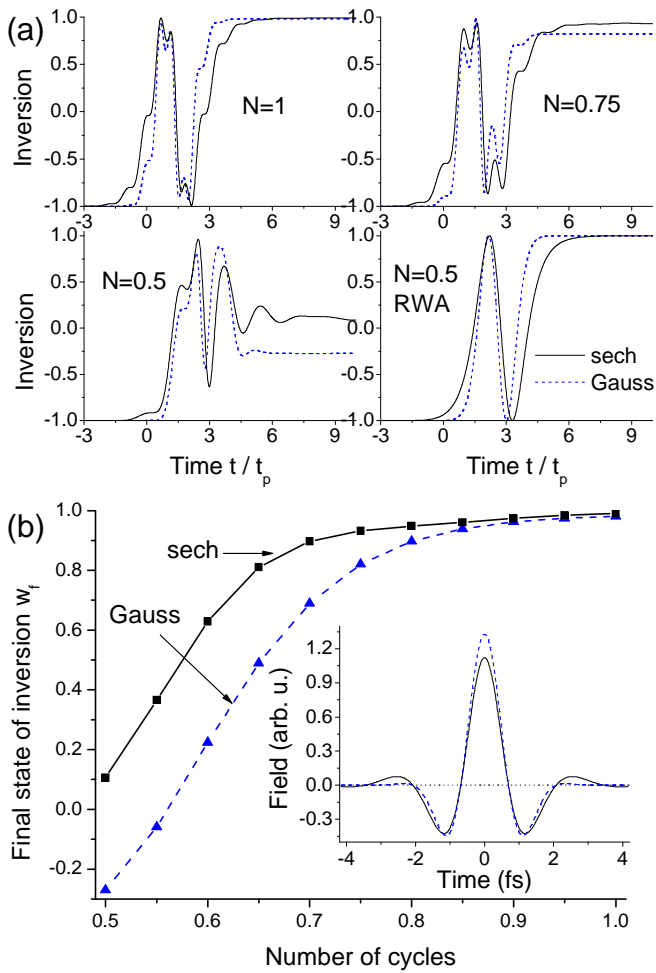}
\caption{\label{fig3} (Color online) (a) Dynamics of the inversion
at the entrance of the two-level medium excited by the Gaussian and
hyperbolic secant $3 \pi$ pulses of different numbers of cycles $N$.
(b) The corresponding dependence of the final state of inversion
(FSI) $w_f$ on the number of cycles. FSI was determined at the time
point $t=50 t_p$. The inset shows the field profiles of the
half-cycle $3 \pi$ pulses of the Gaussian and $\textrm{sech}$
shapes.}
\end{figure}

\begin{figure}[t!]
\includegraphics[scale=0.9, clip=]{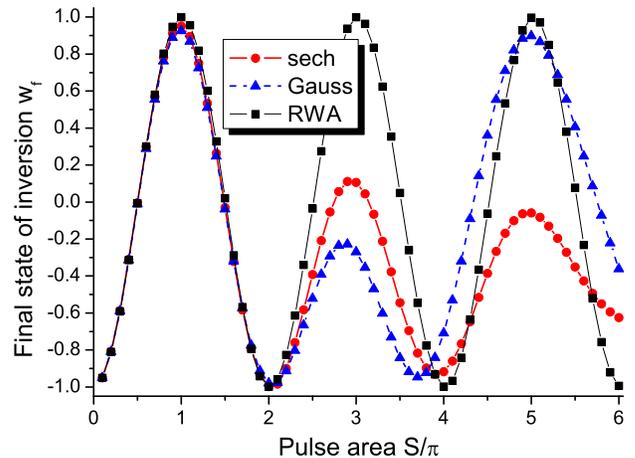}
\caption{\label{fig4} (Color online) The dependence of the final
state of inversion (FSI) $w_f$ on the area of the half-cycle pulses
of the Gaussian and hyperbolic secant shapes. The results for the
RWA case are given for comparison. FSI was determined at the time
point $t=50 t_p$.}
\end{figure}

Let us apply the method described above to simulate propagation of
subcycle pulses in the two-level medium. The main parameter to be
traced is the final state of inversion (FSI) denoted here by $w_f$.
It is the steady value of inversion in which the medium appears
after passage of the incident pulse. The area theorem implies that
FSI depends on the pulse area: if the starting state of inversion is
$w_s=-1$ (two-level system in the ground state), then for the pulse
area $n \pi$ one will have $w_f=1$ or $-1$ at odd and even values of
integer $n$. Figure \ref{fig3}(a) shows the dynamics of inversion at
the entrance of the two-level medium under the influence of the $3
\pi$ subcycle pulse, i.e. for the number of cycles $N \leq 1$. It is
seen that for the single-cycle pulse ($N=1$) the FSI approximately
(though not exactly) corresponds to the value predicted by the area
theorem ($w_f=1$). As the duration of the pulse decreases ($N=0.75$
and $0.5$), the breakdown of the area theorem becomes apparent: such
a short pulse simply cannot guarantee the full cycle of inversion
dynamics, so that $w$ finally appears somewhere in between $-1$ and
$1$. The concrete value of $w_f$ strongly depends on the pulse
shape. We considered two different variants -- secant hyperbolic and
Gaussian pulses -- and see that the difference between the FSI in
these two variants grows as the pulses become shorter. For
comparison, we also calculated the RWA curves for the half-cycle
pulses [see the lower right panel in Fig. \ref{fig3}(a)]. In this
case the area theorem is strictly valid for the pulses of any shape.
In other words, the importance of the rapidly rotating terms in the
Bloch equations increases for ultrashort pulses and results not only
in the breakdown of the area theorem, but also in the strong
dependence of the medium dynamics on the pulse form. This is seen in
the dependence of $w_f$ on the number of cycles $N$ shown in Fig.
\ref{fig3}(b): for the Gaussian subcycle pulses, $w_f$ deviates
faster from the area theorem than in the case of $\textrm{sech}$
pulses. This result seems to be rather surprising, since the field
profiles of $\textrm{sech}$ and Gaussian half-cycle pulses seem to
be not so much different (see the inset) to lead to such a large
difference in $w_f$ ($\Delta w_f \approx 0.35$). Thus, in the range
of subcycle pulses, even a small distinction in the pulse shape may
result in a strong change of the medium dynamics. This should be
taken into account when performing experiments with subcycle pulses
and gives an additional parameter to control the state of the
medium.

The next issue is the dependence of the above described breakdown of
the area theorem on the pulse area. Let us calculate the FSI for
different areas of the incident half-cycle pulse ($N=0.5$). The
results for the Gaussian and hyperbolic secant forms are presented
in Fig. \ref{fig4} as well as the RWA data which are the same for
both pulse profiles. It is seen that at small areas all three curves
approximately coincide and only small deviations from the area
theorem occur as was reported previously \cite{Tarasishin}. But at
the areas above $2 \pi$ the curves rapidly diverge, so that the
large-area half-cycle pulses carry the medium to the state with the
level population which depends on the shape of the pulse. Moreover,
the positions of minima and maxima of inversion also shift from the
usual values. This is especially clearly seen for the second
minimum: while it is the area of $4 \pi$ in the RWA case, it is
significantly lower in the general case -- about $3.9 \pi$ for the
hyperbolic secant pulse and even $3.7 \pi$ for the Gaussian one. It
is also worth noting that the curve for the $\textrm{sech}$ shape is
closer to the RWA dependence in the vicinity of $3 \pi$ areas, but
the situation turns out to be opposite in the vicinity of $5 \pi$
where the curve for the Gaussian pulse approaches the values
consistent with the area theorem. The need of large areas ($\geq 2
\pi$) to observe the strong deviations from the area theorem can be
explained by the fact that the medium affected by the subcycle pulse
does not have enough time to develop the full cycle of dynamics when
it includes more than a single excitation and deexcitation.

\section{\label{concl}Conclusion}

In summary, we have implemented and tested the numerical approach
which allows one to solve the Maxwell--Bloch equations beyond the
RWA and SVEA. This method was used to study the effects of the
interaction of the subcycle pulses with the two-level medium. For
such pulses containing less than one period of optical field, we
have found that (i) the breakdown of the area theorem becomes
apparent at areas larger than $2 \pi$, and (ii) the result of
light-matter interaction strongly depends on the pulse shape. Our
research was focused on the study of the medium dynamics so that we
have not considered the effects of subcycle pulse profile change
such as the formation of an asymmetric wave form \cite{Tarasishin}.
Another interesting issue worth to be studying further is the
possibility or impossibility of subcycle self-induced-transparency
(SIT) solitons. On the one hand, there are studies that show the
existence of few-cycle SIT solitons \cite{Lin}. On the other hand,
the attempts to find a solitonic regime for subcycle pulses were not
successful so far \cite{Tarasishin}. The approach reported in this
paper can be used for a detailed study of this question.

\end{document}